\documentclass{elsart}
\usepackage{natbib}
\usepackage{graphicx}

\begin{document}

\runauthor{Giorgio Riccobene}

\begin{frontmatter}

\title{
Simultaneous measurements of water optical properties by {\it AC9}
transmissometer and {\it ASP-15} Inherent Optical Properties meter
in Lake Baikal}

\author[INR]{V.Balkanov},
\author[JINR]{I.Belolaptikov},
\author[INR]{L.Bezrukov},
\author[IGU]{N.Budnev},
\author[UniRoma1,INFNRoma1]{A. Capone},
\author[IGU]{A.Chensky},
\author[INR]{I.Danilchenko},
\author[INR]{G.Domogatsky},
\author[INR]{Zh.-A.Dzhilkibaev},
\author[NIZH]{S.Fialkovsky},
\author[INR]{O.Gaponenko},
\author[IGU]{O.Gress},
\author[IGU]{T.Gress},
\author[INR]{R.Il'yasov},
\author[INR]{A.Klabukov},
\author[KUR]{A.Klimov},
\author[INR]{S.Klimushin},
\author[INR]{K.Konischev},
\author[INR]{A.Koshechkin},
\author[INR]{Vy.Kuznetzov},
\author[MSU]{L.Kuzmichev},
\author[NIZH]{V.Kulepov},
\author[INR]{B.Lubsandorzhiev},
\author[UniRoma1]{R. Masullo},
\author[INFNLNS,UniCT]{E. Migneco},
\author[INR]{S.Mikheyev},
\author[NIZH]{M.Milenin},
\author[IGU]{R.Mirgazov},
\author[MSU]{N.Moseiko},
\author[MSU]{E.Osipova},
\author[INR]{A.Panfilov},
\author[IGU]{L.Pan'kov},
\author[IGU]{Yu.Parfenov},
\author[IGU]{A.Pavlov},
\author[INFNRoma1]{M. Petruccetti},
\author[JINR]{E.Pliskovsky},
\author[INR]{P.Pokhil},
\author[INR]{V.Poleschuk},
\author[MSU]{E.Popova},
\author[MSU]{V.Prosin},
\author[INFNLNS]{G. Riccobene},
\author[LEN]{M.Rozanov},
\author[IGU]{V.Rubtzov},
\author[IGU]{Yu.Semeney},
\author[DESY]{Ch.Spiering},
\author[DESY]{O.Streicher},
\author[IGU]{B.Tarashansky},
\author[INR]{R.Vasiljev},
\author[DESY]{R.Wischnewski},
\author[MSU]{I.Yashin},
\author[INR]{V.Zhukov}

\address[INR]{Institute for Nuclear Research, 60$^{th}$ Anniversary prospect
7a, 117312, Moscow, Russia}
\address[IGU]{Irkutsk State University, K. Marx street 3, 664003 Irkutsk,
Russia}
\address[MSU]{Skobeltsyn Institute of Nuclear Physics, Moscow State
University, Moscow, Russia}
\address[DESY]{DESY-Zeuthen, Zeuthen, Germany}
\address[JINR]{Joint Institute for Nuclear Research, Dubna, Russia}
\address[NIZH]{Nizhni Novgorod State Technical University, Nizhni Novgorod,
Russia}
\address[LEN]{St. Peterburg State Marine University, St. Peterburg, Russia}
\address[KUR]{Kurchatov Institute, Moscow, Russia}
\address[UniRoma1]{Dipartimento di Fisica Universit\'a La Sapienza, P.le A.
Moro 2, 00185, Roma, Italy}
\address[INFNRoma1]{INFN Sezione Roma-1, P.le A. Moro 2, 00185, Roma, Italy}
\address[INFNLNS]{Laboratori Nazionali del Sud INFN, Via S.Sofia 44, 95123,
Catania, Italy}
\address[UniCT]{Dipartimento di Fisica e Astronomia Universit\'a di Catania,
Corso Italia 57,
95129, Catania, Italy}

\begin{abstract}

Measurements of optical properties in media enclosing \v{C}erenkov
neutrino telescopes are important not only at the moment of the
selection of an adequate site, but also for the continuous
characterization of the medium as a function of time. Over the two
last decades, the Baikal collaboration has been measuring the
optical properties of the deep water in Lake Baikal (Siberia)
where, since April 1998, the neutrino telescope NT-200 is in
operation. Measurements have been made with custom devices. The
NEMO Collaboration, aiming at the construction of a km$^3$
\v{C}erenkov neutrino detector in the Mediterranean Sea, has
developed an experimental setup for the measurement of
oceanographic and optical properties of deep sea water. This setup
is based on a commercial transmissometer. During a joint campaign
of the two collaborations in March and April 2001, light
absorption, scattering and attenuation in water have been
measured. The results are compatible with previous ones reported
by the Baikal Collaboration and show convincing agreement between
the two experimental techniques.

\end{abstract}


\end{frontmatter}

\section{Introduction}

After a long period of experimental work, large \v{C}erenkov
detectors for high energy neutrinos are going to open a new
observational window to the sky. Their main goal is to extend the
volume of the explored Universe by neutrinos, to obtain a
complementary view of astronomical objects and to learn about the
origin of high energy cosmic rays. They are the successors of
underground neutrino detectors which have turned out to be too
small to detect the faint fluxes of neutrinos from cosmic
accelerators.

The new detectors are large, expandable arrays of photomultipliers
constructed in open water or ice. The photomultipliers span a
three-dimensional coarse grid and map the \v{C}erenkov light of
secondary particles produced in  neutrino interactions. Actually,
the basic idea for this detection method goes back to the early
60's \cite{Markov1961}. Pioneering attempts towards its
realization have been made in the course of the DUMAND project
\cite{dumand}. In 1993, the Baikal Collaboration
\cite{Astroparticle-97} succeeded to built the first deep
underwater \v{C}erenkov neutrino detector, which has been stepwise
upgraded to its present stage, NT-200. The AMANDA Collaboration
\cite{Andres2000} has built a \v{C}erenkov detector in the South
Pole ice. Other collaborations (ANTARES \cite{ANTARES}, NESTOR
\cite{Resvanis1993}) are constructing underwater neutrino
detectors of similar size. Since a few years, the NEMO
Collaboration \cite{Capone1999} is performing an intensive R\&D
program  aiming at the construction of a km$^3$ \v{C}erenkov
neutrino telescope in the Mediterranean Sea. Another cubic
kilometer detector, IceCube \cite{Spiering} is planned at the
South Pole. The cubic kilometer scale is set by various
predictions on the extremely low fluxes of high energy neutrinos
expected from astrophysical sources.

In underwater \v{C}erenkov neutrino telescopes, water acts not
only as a target but also as radiator of \v{C}erenkov photons
produced by relativistic charged particles. The detection volume,
as well as the angular and energy resolutions strongly depend on
the water transparency.

The transparency of water as a function of photon wavelength
$\lambda$, is described by the so called inherent optical
properties, like the coefficients for absorption $a(\lambda)$, for
scattering $b(\lambda)$, for attenuation $c(\lambda) = a(\lambda)
+ b(\lambda)$, and by the phase scattering function $\beta(\lambda
, \vartheta)$ (also referred to as volume scattering function)
which represents, for a photon, the probability to be diffused at
an angle $\vartheta$ \cite{Mobley1994}. Another parameter commonly
used in literature is the effective scattering coefficient
$b^{eff}(\lambda) =
b(\lambda)(1-\overline{cos(\lambda,\vartheta)})$, where
$\overline{cos(\lambda,\vartheta)} = \int _{0}^{\pi}
cos(\vartheta) \beta(\lambda,\vartheta) d\vartheta / \int_
{0}^{\pi} \beta(\lambda,\vartheta) d\vartheta$ is the average
cosine of the phase scattering function at a given $\lambda$. The
optical properties of natural water have to be measured {\it
in-situ} in order to allow an unbiased knowledge of light
transmission properties in the medium.

The Baikal collaboration has been investigating the fresh water
deep in Lake Baikal since 1980. The inherent optical properties
have been measured with a series of specially designed devices. It
was shown that the water transparency at depths between 900 m and
1200 m is adequate to operate a neutrino telescope. Put into
operation at April 6$^{th}$, 1998, the neutrino telescope NT-200
incorporates a long-term monitoring system which performs
continuous measurements of the water parameters.
This information
serves as input for Monte-Carlo simulations of the detector
response to atmospheric muons which represent a well-known
calibration source for neutrino telescopes. The muon fluxes
measured with NT-200 are in very good agreement with simulation
results. This fact confirms that the custom-made devices and the
methods to extract the relevant information on optical parameters
yield reliable results.

The NEMO collaboration has been investigating oceanographic and
optical properties of several deep sea marine sites close to the
Italian coast, with the aim to select the optimal site for the
construction of a km$^3$ detector in the Mediterranean Sea.
Absorption and attenuation coefficients for light in the
wavelength region between 412$\div$715 nm \cite{Capone2001} have
been measured with a set-up based on commercial devices.

Optical measurements in deep water are extremely difficult, and
possible systematic errors related to these measurements suggest
careful cross checks of results by complementary methods. For
these reasons, during March - April 2001, the NEMO and Baikal
Collaborations have started a joint campaign to measure the
optical properties of deep water in Lake Baikal using two
different devices. One set-up is based on the transmissometer {\it
AC9}, operated by the NEMO group, the other device, {\it ASP-15}
(Absorption, Scattering and Phase function meter), was developed
and operated by the Baikal Collaboration. The cross check of
experimental results has been crucial for both devices, since both
have an excellent sensitivity in measuring water optical
properties, however, they can be affected by different sources of
systematic errors which could deteriorate the absolute accuracy.
The measurements reported in the following sections have been
carried out during March - April 2001, from the ice camp above the
neutrino telescope NT-200.

\section{Instrumentation and data acquisition}

\subsection{The {\it AC9} transmissometer}

The {\it AC9}, manufactured by {\it WETLabs} \cite{Wetlabsmanual},
is a transmissometer capable to measure absorption and attenuation
coefficients at nine different wavelengths in the range $412 \div
715$ nm. Using an accurate calibration procedure, the NEMO
collaboration has achieved an accuracy of about $1.5 \cdot
10^{-3}$ m$^{-1}$ in $a$ and $c$ measurements \cite{Capone2001}.

\begin{figure}[hbtp]
\centerline {\includegraphics[width=8cm]{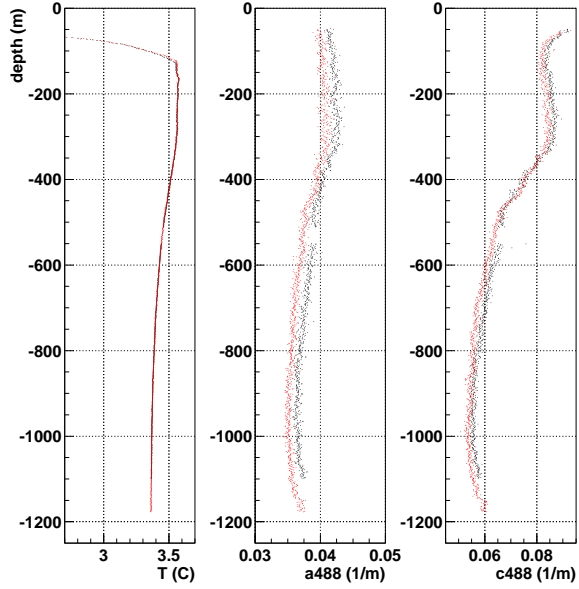}}
\caption{Profiles of water temperature and water optical
properties (absorption and attenuation coefficients: $a(488)$ and
$c(488)$), as functions of depth, obtained by two measurements
(red and black dots) with the {\it AC9}-CTD set-up in Lake Baikal
during March 2001. The NT-200 telescope is located between 1100 m
and 1170 m depth.}\label{fig:Tempacbaikal}
\end{figure}

During the measurements in Lake Baikal, the {\it AC9} device and a
CTD (a probe that measures water conductivity, temperature and
pressure) were operated through an electro-mechanical cable. With
this set-up we have obtained two vertical profiles of the water
column (50 m$<$ depth $<$1100 m), collecting about ten data-sets
per meter of depth. Each data-set consists of a measurement of
temperature and optical properties, $a(\lambda)$ and $c(\lambda)$,
over the nine wavelengths.

In figure \ref{fig:Tempacbaikal} we show, as a function of depth,
the water temperature together with the values for absorption and
attenuation coefficient at $\lambda=488$ nm measured during the
first and the second deployment in Lake Baikal (for discussion see
section \ref{sec:results}).

\subsection{{\it ASP-15} - an instrument for long-term
monitoring of the inherent optical properties of deep water}

\begin{figure}[hbt]
 \centerline
{\includegraphics[width=7cm]{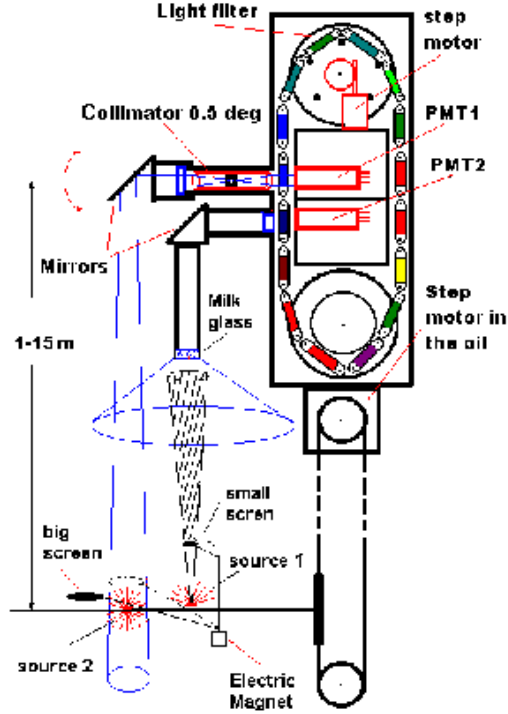}} \caption{The {\it ASP-15}
device  for long-term monitoring of deep water inherent optical
properties.}\label{fig:ASP-15}
\end{figure}

The {\it ASP-15} device (see figure \ref{fig:ASP-15}) has two receiving
channels:
one with a wide aperture to measure $a$ and $b$  and another one
with a rotating mirror and a narrow angle collimator to measure
the phase scattering  function.

Two photomultipliers (type FEU-130) and 15 interference light
filters are assembled in a cylindrical container. The filters
wavelengths are ranged from 369 nm to 691 nm. Both
photomultipliers operate in photon counting mode. Two isotropic
point-like light sources and two screens are assembled on a frame,
which can be moved by a stepping motor over distances ranging from
0.4 to 15 meters with respect to the milk glass window.
Measurements were carried out separately for each source and
controlled via cable by a computer on shore or at the ice camp.
The device is described in detail in
\cite{Tarash-94,Tarash-95,Gapon-97}. The principle of the $a$
measurement with {\it ASP-15} is described in \cite{Bezrukov-90}.
We measure the dependence of the luminosity $E$ on the distance
$R$ between source and receiver with each of the 15 light filters
and approximate the absorption coefficient by

\begin{center}
\begin{equation}
\label{abs}
a=- \frac {ln(E_1 \cdot R_1^2 / E_2 \cdot R_2^2)} {R_1-R_2},
\end{equation}
\end{center}

where $E_1$ and $E_2$ are the luminosity at distances $R_1$ and
$R_2$, respectively. Monte-Carlo simulations \cite{Bezrukov-90}
have shown that for an isotropic or a Lambertian (cosine) light
source in water, the difference between approximation (\ref{abs})
(which is an exact definition of $a$ in a case of a medium without
scattering and isotropic point like source) and the exact value of
$a$ is less than $1\%$, provided a strongly anisotropic phase
scattering function, low scattering and $R \leq a^{-1}$.

The scattering coefficient $b$ is approximated \cite{Zeuthen-98}
by
\begin{center}
\begin{equation}
b=ln(1-E_s/E)/R,
\end{equation}
\end{center}
where $E_s$ and $E$ are the luminosity at distance $R$
from the screened and unscreened source, respectively.

Our estimation of the total uncertainties due to approximations
(1,2) and systematic errors is: $\Delta a(\lambda) \leq$ 5\%  for
$a \geq 0.02$ m$^{-1}$ and $\Delta b(\lambda) \leq$ 10 \% for $b
\geq 0.02$ m$^{-1}$. In all figures below, only statistical errors
are shown for the {\it ASP-15} data.

\section{Results}
\label{sec:results}
\subsection{Light absorption in Lake Baikal}
\label{sec:absorption}

\begin{figure}[hbt]
\centerline {\includegraphics[width=9cm]{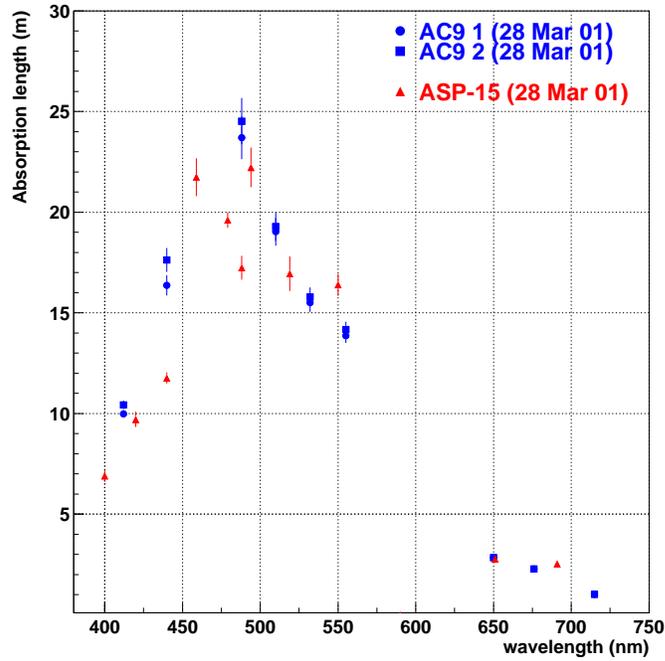}}
\caption{Absorption length measured with {\it
ASP-15} at a depth of 200 m. {\it AC9} data are averaged over the
depth interval 180$\div$220 m.}\label{fig:ablength200}

\end{figure}

In this section we discuss the results of light absorption
measurements performed with both devices {\it AC9} and {\it
ASP-15}.
In table \ref{tab:abcoeff200} and figure \ref{fig:ablength200}
we present respectively the absorption coefficients
and absorption lengths ($L_a(\lambda)=1/a(\lambda)$) as a function of wavelength. {\it
ASP-15}
data  have been taken at a depth of 200 m, the {\it AC9} values
are the average of data collected at depths between 180 m and 220 m.

\begin{table}[tb]
\caption{Absorption coefficients measured during two deployments
of {\it AC9} (March 28$^{th}$) and during one deployment of {\it
ASP-15} (March 28$^{th}$) at a depth of 200 m.{\it AC9} data are averaged over the depth interval
180$\div$220 m.}
\label{tab:abcoeff200}

\begin{center}
\begin{small}
\begin{tabular}{cccc} \hline
  $\lambda$ & {\it AC9} 1 & {\it AC9} 2 & {\it ASP-15} 28/03\\
  (nm) & $a$(m$^{-1}$) & $a$(m$^{-1}$) & $a$(m$^{-1}$) \\
  369   &                     &                    & 0.212$\pm$0.026  \\
  374   &                     &                    & 0.264$\pm$0.006  \\
  400   &                     &                    & 0.145$\pm$0.006  \\
  412   &  0.100$\pm$0.003    &  0.096$\pm$0.003   &                 \\
  420   &                     &                    & 0.103$\pm$0.004  \\
  440   &  0.061$\pm$0.002    &  0.057$\pm$0.002   & 0.085$\pm$0.002  \\
  459   &                     &                    & 0.046$\pm$0.002  \\
  479   &                     &                    & 0.051$\pm$0.001  \\
  488   &  0.042$\pm$0.001    &  0.041$\pm$0.001   & 0.058$\pm$0.002  \\
  494   &                     &                    & 0.045$\pm$0.002  \\
  510   &  0.052$\pm$0.001    &  0.052$\pm$0.001   &                 \\
  519   &                     &                    & 0.059$\pm$0.003  \\
  532   &  0.064$\pm$0.001    &  0.063$\pm$0.001   &                 \\
  550   &                     &                    & 0.061$\pm$0.002  \\
  555   &  0.072$\pm$0.001    &  0.070$\pm$0.001   &                 \\
  650   &  0.352$\pm$0.001    &  0.351$\pm$0.001   &                 \\
  651   &                     &                    & 0.361$\pm$0.006  \\
  676   &  0.439$\pm$0.001    &  0.439$\pm$0.001   &                 \\
  691   &                     &                    & 0.395$\pm$0.012  \\
  715   &  0.979$\pm$0.001    &  0.979$\pm$0.001   &                 \\
\hline
\end{tabular}
\end{small}
\end{center}
\end{table}

\begin{figure}[bt]
 \centerline
{\includegraphics[width=9cm]{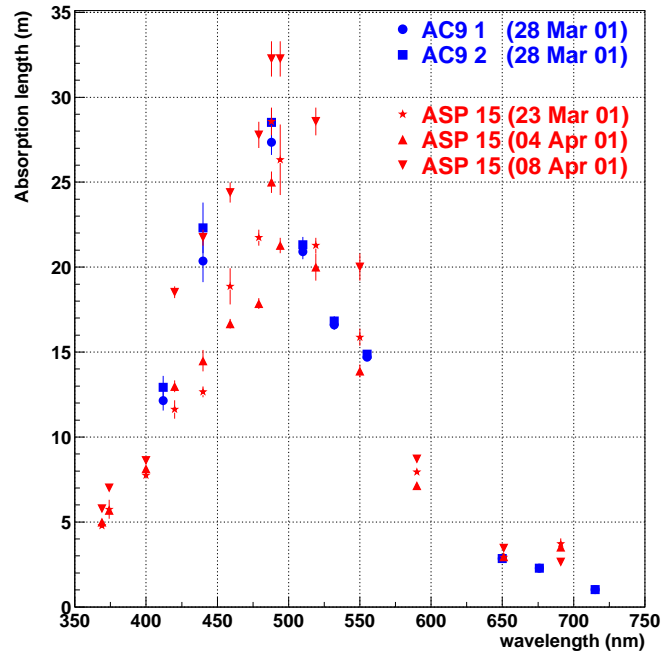}}
\caption{Absorption length measured with {\it
ASP-15} at 1000m depth. {\it AC9} data are averaged over the depth interval
980$\div$1020 m.} \label{fig:ablength1000}
\end{figure}

Table \ref{tab:abcoeff1000} and figure \ref{fig:ablength1000} show
the results obtained for absorption coefficients and absorption
lengths
at a depth of 1000 m ({\it ASP-15}) and for depths between
980 m and 1020 m ({\it AC9}).

\begin{table}[tb]
\caption{Absorption coefficients measured during two deployments
of the {\it AC9} (March 28$^{th}$) and during three deployments of
{\it ASP-15} (March 23$^{rd}$, April 4$^{th}$ and April 8$^{th}$)
in Lake Baikal at 1000 m depth.{\it AC9} data are averaged over the depth interval
980$\div$1020 m.} \label{tab:abcoeff1000}
\begin{center}
\begin{small}
\begin{tabular}{cccccc} \hline
  $\lambda$ & {\it AC9} 1& {\it AC9} 2& {\it ASP-15} 23/03 & {\it ASP-15}
04/04
  & {\it ASP-15} 08/04 \\
  (nm) & $a$ (m$^{-1}$) & $a$ (m$^{-1}$) & $a$ (m$^{-1}$) & $a$ (m$^{-1}$)
  & $a$ (m$^{-1}$) \\
  369   &                     &                    & 0.209$\pm$0.007  &
0.200$\pm$0.004 &   0.173$\pm$0.006  \\
  374   &                     &                    & 0.174$\pm$0.017  &
0.176$\pm$0.004 &   0.143$\pm$0.004  \\
  400   &                     &                    & 0.129$\pm$0.003  &
0.123$\pm$0.003 &   0.116$\pm$0.003  \\
  412   &  0.082$\pm$0.003    &  0.077$\pm$0.003   &                &
&                  \\
  420   &                     &                    & 0.086$\pm$0.004  &
0.077$\pm$0.002 &   0.054$\pm$0.001  \\
  440   &  0.049$\pm$0.002    &  0.045$\pm$0.002   & 0.079$\pm$0.002  &
0.069$\pm$0.003 &   0.046$\pm$0.001  \\
  459   &                     &                    & 0.053$\pm$0.003  &
0.060$\pm$0.001 &   0.041$\pm$0.001  \\
  479   &                     &                    & 0.046$\pm$0.001  &
0.056$\pm$0.001 &   0.036$\pm$0.001  \\
  488   &  0.037$\pm$0.001    &  0.035$\pm$0.001   & 0.035$\pm$0.001  &
0.040$\pm$0.001 &   0.031$\pm$0.001  \\
  494   &                     &                    & 0.038$\pm$0.003  &
0.047$\pm$0.001 &   0.031$\pm$0.001  \\
  510   &  0.048$\pm$0.0015   &  0.047$\pm$0.001   &                &
&                  \\
  519   &                     &                    & 0.047$\pm$0.001  &
0.050$\pm$0.002 &   0.035$\pm$0.001  \\
  532   &  0.060$\pm$0.001    &  0.059$\pm$0.001   &                &
&                  \\
  550   &                     &                    & 0.063$\pm$0.002  &
0.072$\pm$0.002 &   0.050$\pm$0.002  \\
  555   &  0.068$\pm$0.001    &  0.067$\pm$0.001   &                &
&                  \\
  590   &                     &                    & 0.126$\pm$0.003  &
0.140$\pm$0.003 &   0.115$\pm$0.003  \\
  650   &  0.351$\pm$0.001    &  0.351$\pm$0.001   &                &
&                  \\
  651   &                     &                    & 0.343$\pm$0.003  &
0.338$\pm$0.008 &   0.290$\pm$0.006  \\
  676   &  0.439$\pm$0.001    &  0.439$\pm$0.001   &                &
&                  \\
  691   &                     &                    & 0.269$\pm$0.023  &
0.284$\pm$0.008 &   0.379$\pm$0.021  \\
  715   &  0.984$\pm$0.001    &  0.984$\pm$0.001   &                &
&                  \\
\hline
\end{tabular}
\end{small}
\end{center}
\end{table}

The two sets of {\it AC9} data were collected with about ten hours
time difference. Each measurement was preceded by an accurate
cleaning of the instrument optics and by a calibration. The
agreement between the results obtained from the two data sets
confirms the reliability of the calibration procedure.

The agreement of the results obtained by means of {\it AC9} and
{\it ASP-15} at 1000 m depth is rather good: in spite of the fact
that the two instruments are based on different methodologies and
have different sources of systematic errors, the central values
are compatible. This result proves the validity of both
measurement techniques.

The spread between data collected with {\it AC9} and {\it ASP-15}
at 200 m depth can be attributed to local changes in optical and
hydro-physical properties of the water column, extensively
discussed by the Baikal Collaboration in \cite{Ocean Phys1,Ocean
Phys2}. The two data samples have been collected at two sites with
about 100 m distance.

At a depth of 200 m, the maximum value of the absorption length is located
in the blue-green region, at $\lambda \sim 490$ nm. The average
of the measured values are $L_a = 24.1 \pm 0.5 $ m  for  {\it AC9}
($\lambda =$ 488 nm) and $L_a = 22.2 \pm 1.0$ m  for {\it ASP-15}
($\lambda =$ 494 nm).

For the data samples collected at 1000 m depth, the maximum values
of the absorption lengths are also observed at $\lambda \sim 490$
nm. Their mean values are: $L_a = 27.9 \pm 0.7 $ m for {\it AC9}
($\lambda =$ 488 nm) and  $L_a = 28.3 \pm 1.5$ m  for {\it ASP-15}
($\lambda =$ 488 nm ). These values
do not contradict previous measurements of the Baikal Collaboration
\cite{Bezrukov-90,Zeuthen-98,Laser-99}.

The obvious differences between optical properties at 1000 m and
200 m are due to the different characteristics of Lake Baikal
waters below and above the boundary depth of solar radiation
penetration, which is located at a depth of about $\sim$ 400 m
(see figure \ref{fig:Tempacbaikal}). Above the solar radiation
boundary depth the water column shows a time dependent behavior,
strongly influenced by biological activity. Below the solar
radiation boundary depth, where the water column is more stable
and the biological activity is reduced, the water transparency
increases (about 25$\%$ increase of $L_a$ at blue wavelengths).
The best water transparency is measured between 900 m and 1150 m,
covering the vertical extension of the NT-200 telescope. Below the
absorption length decreases, probably due to the water streamed
along the very steep slope of the lake bed.

\subsection {Light attenuation and scattering in Lake Baikal}
\label{sec:scattering}

While {\it ASP-15} is designed to measure directly
the absorption, $a(\lambda)$,
and scattering, $b(\lambda)$, coefficients the
{\it AC9} measures
the absorption, $a(\lambda)$, and
attenuation, $c(\lambda)$ coefficients.
In the latter case the
scattering coefficient can be obtained as the
difference between
absorption and attenuation coefficients ($b(\lambda)=c(\lambda) - a(\lambda)$)
and compared to
the results from the direct measurements with {\it ASP-15}.

In tables \ref{tab:atcoeff200} and \ref{tab:atcoeff1000} we
present the attenuation coefficients measured at
depths of about 200 m and 1000 m with {\it AC9}.

\begin{table}[htb]
\caption{Mean attenuation coefficients measured during two
deployments of the {\it AC9} (March $28^{th}$) at depths between 180 m and 220 m}
\label{tab:atcoeff200}
\begin{center}
\begin{small}
\begin{tabular}{ccc} \hline
  $\lambda$ & {\it AC9} 1 & {\it AC9} 2 \\
  (nm) & $c$ (m$^{-1}$) & $c$ (m$^{-1}$) \\
  412   &  0.162$\pm$0.002  &  0.160$\pm$0.002  \\
  440   &  0.118$\pm$0.002  &  0.116$\pm$0.002  \\
  488   &  0.086$\pm$0.001  &  0.084$\pm$0.001 \\
  510   &  0.094$\pm$0.001  &  0.093$\pm$0.001 \\
  532   &  0.101$\pm$0.001  &  0.100$\pm$0.001  \\
  555   &  0.108$\pm$0.001  &  0.107$\pm$0.001  \\
  650   &  0.391$\pm$0.002  &  0.389$\pm$0.002  \\
  676   &  0.476$\pm$0.002  &  0.472$\pm$0.002  \\
  715   &  1.015$\pm$0.001  &  1.012$\pm$0.001   \\
\hline
\end{tabular}
\end{small}
\end{center}
\end{table}

\begin{table}[tb]
\caption{Mean attenuation coefficients measured during two
deployments of the {\it AC9} (March $28^{th}$) at depths between 980 m and 1020 m.}
\label{tab:atcoeff1000}
\begin{center}
\begin{small}
\begin{tabular}{ccc} \hline
  $\lambda$  & {\it AC9} 1& {\it AC9} 2\\
  (nm)  &  $c$ (m$^{-1}$) & $c$ (m$^{-1}$) \\
  412   &  0.123$\pm$0.002   &  0.120$\pm$0.002  \\
  440   &  0.085$\pm$0.002   &  0.082$\pm$0.002  \\
  488   &  0.056$\pm$0.001   &  0.053$\pm$0.001  \\
  510   &  0.065$\pm$0.001   &  0.064$\pm$0.001  \\
  532   &  0.072$\pm$0.001   &  0.070$\pm$0.001  \\
  555   &  0.090$\pm$0.001   &  0.088$\pm$0.001  \\
  650   &  0.373$\pm$0.002   &  0.370$\pm$0.002  \\
  676   &  0.455$\pm$0.002   &  0.451$\pm$0.002  \\
  715   &  0.997$\pm$0.001   &  0.995$\pm$0.001  \\
\hline
\end{tabular}
\end{small}
\end{center}
\end{table}

Tables \ref{tab:scatt200} and \ref{tab:scatt1000} show
 the values of the scattering coefficients
measured with {\it ASP-15} at depths of 200 m and 1000 m,
respectively.

\begin{table}
\caption{Scattering coefficients measured with {\it ASP-15} at
a depth of 200 m (March $27^{th}$).}
\label{tab:scatt200}
\begin{center}
\begin{small}
\begin{tabular}{cc} \hline
  $\lambda$ & {\it ASP-15} 27/03 \\
  (nm)    & $b$ (m$^{-1}$)  \\
  400   & 0.039$\pm$0.004  \\
  420   & 0.035$\pm$0.002  \\
  440   & 0.034$\pm$0.002  \\
  459   & 0.035$\pm$0.009  \\
  479   & 0.033$\pm$0.001  \\
  488   & 0.033$\pm$0.002  \\
  494   & 0.033$\pm$0.002  \\
  519   & 0.032$\pm$0.001  \\
  550   & 0.031$\pm$0.001  \\
  590   & 0.029$\pm$0.001  \\
  651   & 0.026$\pm$0.005  \\
\hline
\end{tabular}
\end{small}
\end{center}
\end{table}

\begin{table}
\caption{Scattering coefficients measured with {\it ASP-15} at
a depth of 1000 m (April $04^{th}$ and April $08^{th}$).}
\label{tab:scatt1000}
\begin{center}
\begin{small}
\begin{tabular}{ccc} \hline
  $\lambda$ & {\it ASP-15} 04/04 & {\it ASP-15} 08/04\\
  (nm) & $b$ (m$^{-1}$) & $b$ (m$^{-1}$) \\
  369   &                   &   0.145$\pm$0.005   \\
  374   &                   &   0.155$\pm$0.005   \\
  400   &  0.033$\pm$0.005  &   0.047$\pm$0.005   \\
  420   &  0.030$\pm$0.005  &   0.044$\pm$0.005   \\
  440   &  0.022$\pm$0.002  &   0.035$\pm$0.002   \\
  459   &  0.016$\pm$0.002  &   0.023$\pm$0.002   \\
  479   &  0.015$\pm$0.001  &   0.022$\pm$0.001   \\
  488   &  0.014$\pm$0.001  &   0.020$\pm$0.001   \\
  494   &  0.014$\pm$0.001  &   0.021$\pm$0.001   \\
  519   &  0.014$\pm$0.001  &   0.021$\pm$0.001   \\
  550   &  0.013$\pm$0.001  &   0.016$\pm$0.001   \\
  590   &  0.015$\pm$0.006  &   0.024$\pm$0.006   \\
  651   &                   &   0.095$\pm$0.006   \\
\hline
\end{tabular}
\end{small}
\end{center}
\end{table}

\begin{figure}[bt]
\centerline {\includegraphics[width=9cm]{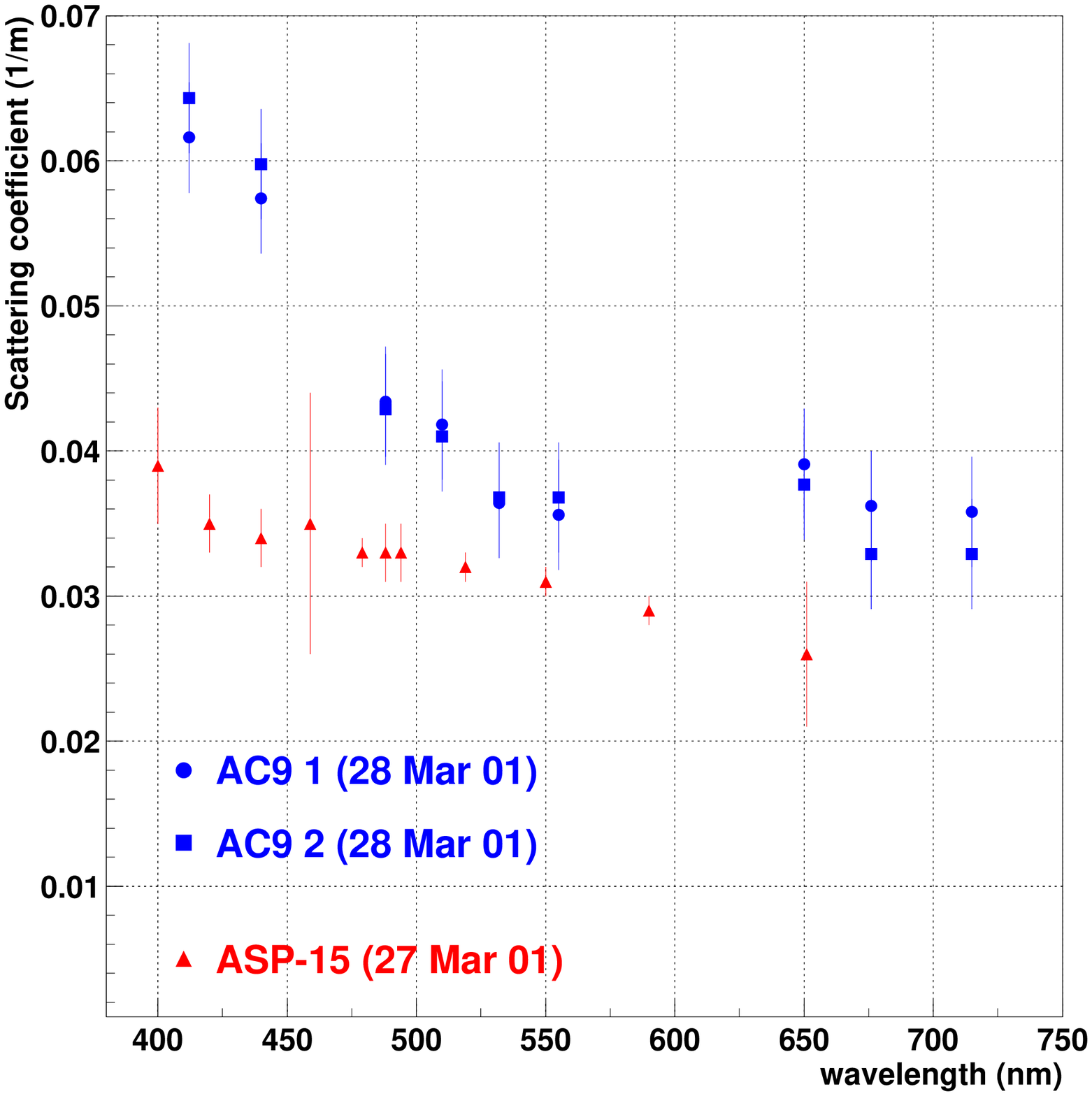}}
\caption{Scattering coefficients estimated from {\it AC9} data and
measured with {\it ASP-15} at a depth of 200
m.}\label{fig:scatt200}
\end{figure}

\begin{figure}[bt]
\centerline {\includegraphics[width=9cm]{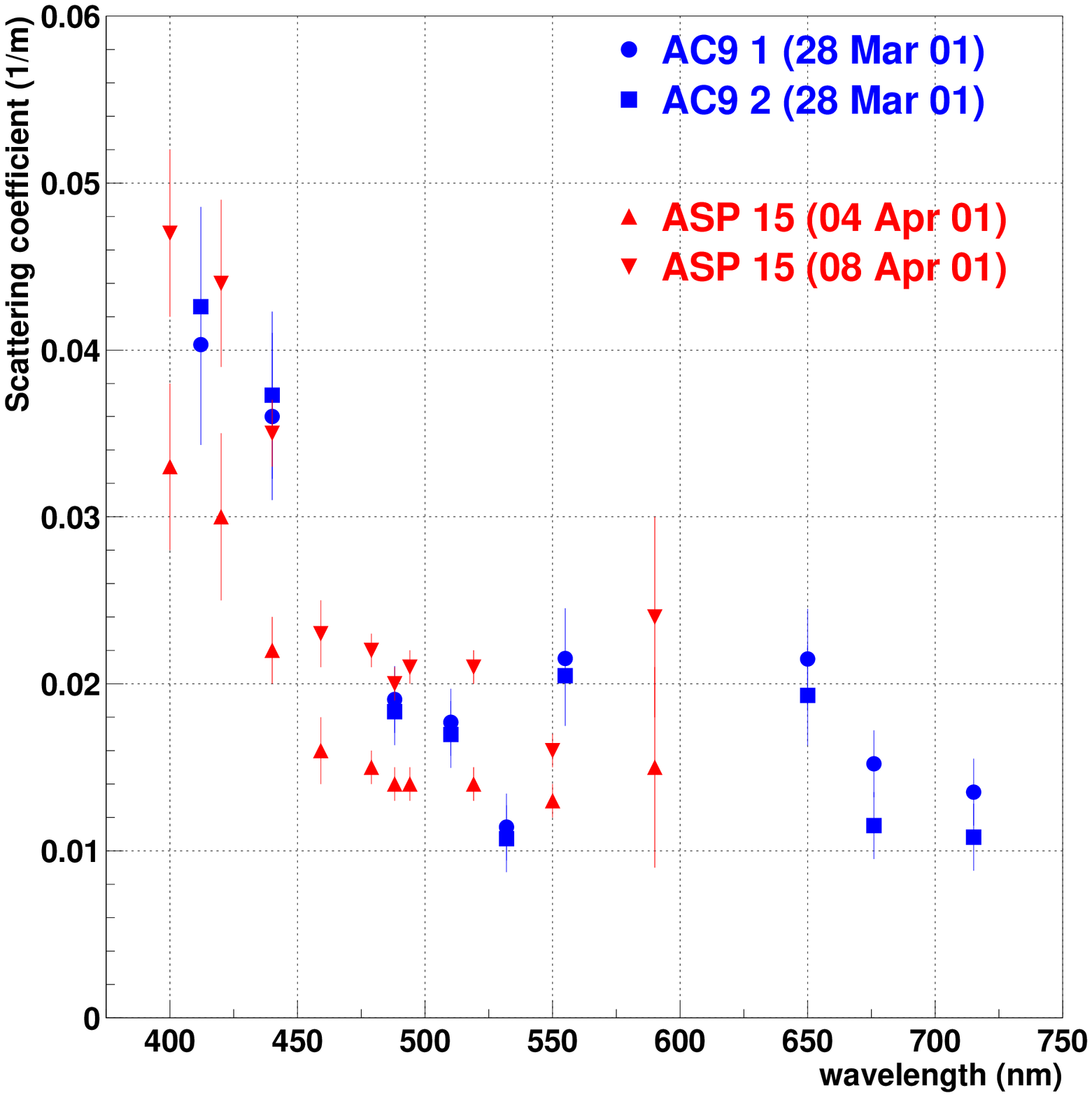}}
\caption{Scattering coefficients estimated from {\it AC9} data and
measured with {\it ASP-15} at a depth of 1000
m.}\label{fig:scatt1000}

\end{figure}

Figures \ref{fig:scatt200} and \ref{fig:scatt1000} show the
comparison between the scattering coefficients, $b(\lambda)$,
measured directly by {\it ASP-15} (tables \ref{tab:scatt200} and
\ref{tab:scatt1000})and evaluated from the
absorption and attenuation coefficients measured by {\it AC9} :
($a(\lambda)$
from tables \ref{tab:abcoeff200},\ref{tab:abcoeff1000} and
$c(\lambda)$ from tables
\ref{tab:atcoeff200},\ref{tab:atcoeff1000}).

Figure \ref{fig:scatt1000} shows good agreement between results
obtained  with {\it AC9} and {\it ASP-15} at 1000 m depth. At 200
m depth (figure \ref{fig:scatt200}) there are discrepancies
which confirm the different optical properties of the water layers
measured by {\it AC9} and {\it ASP-15}, already indicated by the
results of the absorption measurements at the same depth (see
section \ref{sec:absorption}).

Given the strong water currents
at shallow depth and inhomogeneous distribution of biologically
active substances, a strong variation of
optical parameters within one day appears to be realistic.

\begin{figure}[bt]
\centerline {\includegraphics[width=9cm]{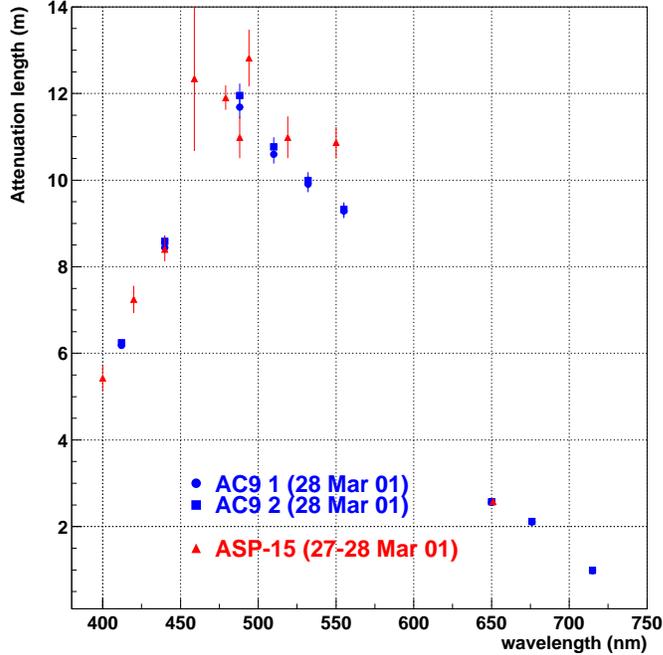}}
\caption{Attenuation length measured with {\it AC9} (March $28^{th}$) and {\it
ASP-15} at 200 m depth. The values of $c(\lambda)$ for {\it
ASP-15} are the sum of the values of $a(\lambda)$ measured in
March 28$^{th}$ and the values of $b(\lambda)$ measured in March
27$^{th}$.}\label{fig:atlength200}
\end{figure}

\begin{figure}[bt]
\centerline {\includegraphics[width=9cm]{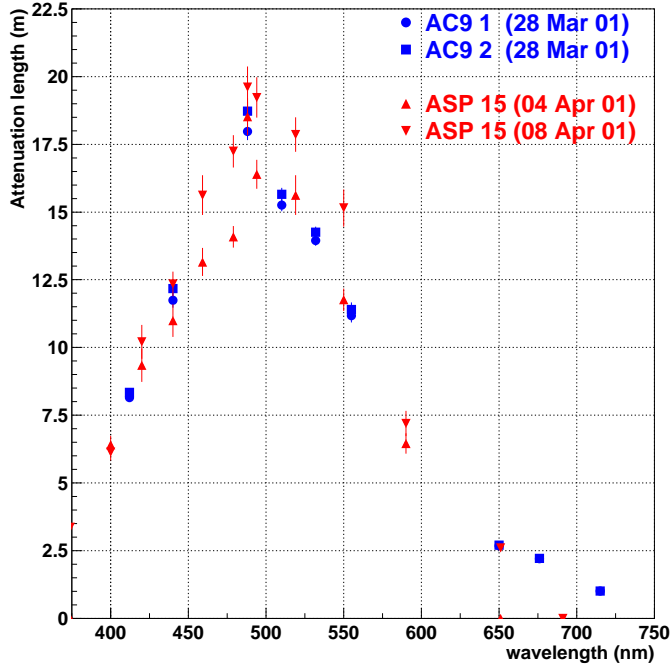}}
\caption{Attenuation length measured with {\it AC9} (March $28^{th}$) and {\it
ASP-15} at 1000 m depth. The values $c(\lambda)$ for {\it
ASP-15} are the sum of $a(\lambda)$ and the values of $b(\lambda)$
measured during two runs on April 4$^{th}$ and on April
8$^{th}$.}\label{fig:atlength1000}

\end{figure}

At last we show in figures \ref{fig:atlength200} and
\ref{fig:atlength1000} the values of the attenuation lengths
($L_c(\lambda)=1/c(\lambda)$) obtained at depths of 200 m and 1000
m. The values of $c(\lambda)$ for {\it ASP-15} are obtained adding
the absorption and scattering coefficients reported in tables
\ref{tab:abcoeff200},\ref{tab:abcoeff1000},\ref{tab:scatt200},\ref{tab:scatt1000},
while for {\it AC9} they are measured directly (see tables
\ref{tab:atcoeff200},\ref{tab:atcoeff1000}). To evaluate the {\it
ASP-15} results at 200 m depth we have used the absorption data
measured in March 28$^{th}$ and the scattering data measured on
March 27$^{th}$. Figure \ref{fig:atlength1000} shows good
agreement between results from {\it AC9} and {\it ASP-15}.
Similarly to the absorption coefficient, the attenuation
coefficient has its smallest value in the region of $\lambda \sim$
490 nm for both depths.

\section{Conclusion}

Measurements of the optical water properties in Lake Baikal
confirm that the NT-200 telescope is located at optimal depth,
where light absorption and attenuation processes are the smallest.
Data have been collected with two instruments, which use different
measurement principles and have different sources of systematic
errors. Data show that, at a depth of 1000 m, the highest
transparency is observed for $\lambda=$ 488 nm. The measured
values for absorption length $L_a$, scattering length $L_b$ and
attenuation length $L_c$ at 1000 m depth are: $L_a(488)=27.9 \pm
0.7$ m, $L_c(488)=18.3 \pm 0.3$ m as measured with {\it AC9} and
$L_a(488)=28.3 \pm 1.0$ m, $L_b(488)=58.8 \pm 3.5$ m as measured
with {\it ASP-15}. The depth profile of the absorption coefficient
measured by {\it AC9} (see figure \ref{fig:Tempacbaikal}) shows
the effect of biologically active substances and mineral
particulate suspended in water. This effect is very conspicuous in
the depth range 0 $\div$ 400 m (above the boundary depth of
penetration of solar radiation), and
starts to be visible again for depth higher than 1150 m, near the lake bed.\\
The obtained results demonstrate that the systematic errors are
rather small for both instruments and validate the use of both
devices to characterize {\it in situ} the inherent optical
properties of underwater sites.

\section{Acknowledgments}
This work was supported by the Russian Ministry of Industry,
Science and Technology (contract 102-11(00)-p), the Russian
Ministry of Education, the German Ministry of Education and
Research, Russian Fund of Basic Research (grants 99-02-1837a,
01-02-31013 and 00-15-96794), the Russian Federal Program {\it
Integration}, the Program {\it Universities of Russia} and UNESCO
Chair of Water Researches and
the Italian Istituto Nazionale di Fisica Nucleare (INFN).

\end{document}